\documentclass[aps,prb]{revtex4-1}

\usepackage{graphicx}

\begin{document}

\title{Use of a sigmoid function to describe second peak in magnetization loops}
\author{Denis~Gokhfeld}
\affiliation{Kirensky Institute of Physics, Federal Research Center KSC SB RAS,
      Akademgorodok 50, Krasnoyarsk, 660036 Russia.}
\date{\today}

\begin{abstract}
Order-disorder transitions of a vortex lattice transfer type-II superconductors from a low critical current state to a high one. The similar transition between different current states can be caused by electromagnetic granularity.
A sigmoid curve is proposed to describe the corresponding peak in a field dependence of the macroscopic critical density.
Using the extended critical state model, analytic expressions are obtained for the field dependencies of the local critical current density, the depth of equilibrium surface region, and the macroscopic critical current density. The expressions are well fit to published data.

\end{abstract}


\maketitle

\section{Introduction}
Some superconducting samples have magnetization loops with a second peak (fishtail peculiarity) in high magnetic fields. 
Reasons of the peak effect are generally attributed to a phase transition of vortex lattice \cite{Vinokur, Ann, Babich, Zehetmayer} or a magnetic phase separation  \cite{Gorb90,Nagaev95,Kenzelmann17}.

\begin{figure}[ht]
\centering
\includegraphics[width=4in]{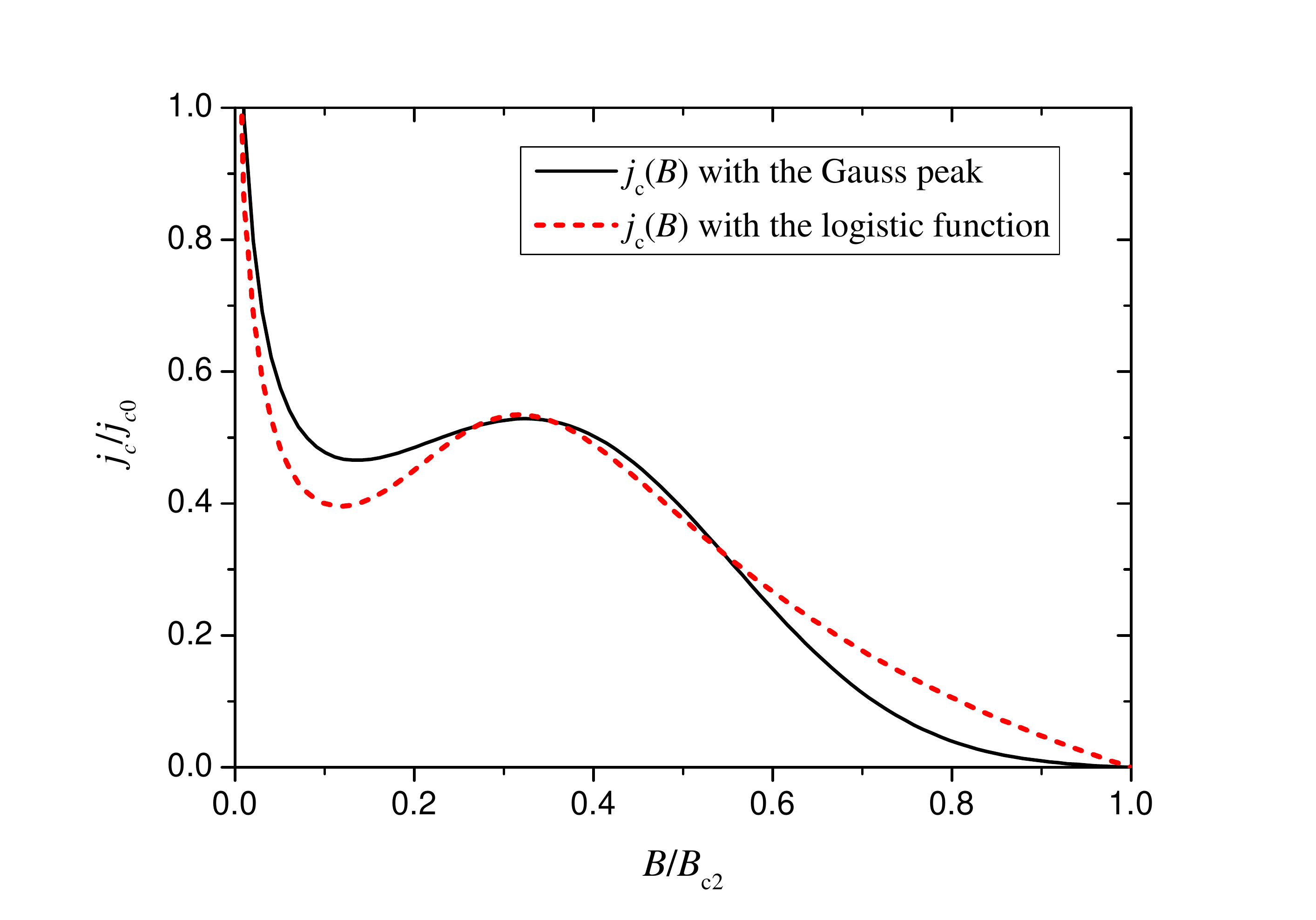}
\caption{Comparison of the Gauss function and the logistic function as a source of the peak in $j_c^{\rm peak}(B)$ dependencies.}
\label{fig_jcb}
\end{figure}

Earlier, the critical state model \cite{Johansen97,Chandran98,Chandran99,Ravikumar99,Inanir07} and the extended critical state model \cite{gokhfeld_peak,gokhfeld_jetp} were used to describe the peak effect without considering underlying mechanisms. 
In these works nonmonotonic dependencies of the critical current density $j_c$ on magnetic field $B$ are suggested. The depth of equilibrium surface region $l_s$ shrinks at the magnetic field range corresponding to the peak effect that is accounted by the extended critical state model. To obtain the peak at high fields a bell-shaped function $f_{\rm peak}(B)$ is added to a monotonic decreasing function $j_c(B)$. The $f_{\rm peak}(B)$ function provides the growing part of the resulted nonmonotonic $j_c^{\rm peak}(B)$ function at high fields. The decreasing part of the $f_{\rm peak}(B)$ function is not important to reveal the peak because the unperturbed $j_c(B)$ dependence provides fast decrease of the $j_c^{\rm peak}(B)$ dependence at high fields. 
A nondecreasing function, e.g. the logistic function, can be used as the $f_{\rm peak}(B)$ function to provide the peak. Figure~\ref{fig_jcb} demonstrates that the curve computed with a bell-shaped $f_{\rm peak}(B)$ function (the Gauss function) and the curve computed with the logistic function are quite similar.  

The logistic function is a part of a Boltzmann sigmoid function, which is typically used to describe crossovers between phases \cite{Zablotskii,Sukhareva,Navarro,Derev17}. The Boltzmann sigmoid function is written as
\begin{equation}
Y(t)=Y_1+\frac{Y_2-Y_1}{1+e^{-({t-t_{tr}})/{t_w}}}~~,\\
\label{e1}
\end{equation}
where $Y_1$ and $Y_2$ are some quantities characterizing correspondingly two different phases, $t$ is a variable, $t_{tr}$ is the transition middle and $t_w$ is the transition wide. 
In next section we apply the Boltzmann sigmoid function to describe the peak 	n the field dependence of the critical current density.


\section{Peak effect and critical current}


The order-disorder transition of the 2D vortex lattice results in an increasing of the local critical current density $j_c$ \cite{Vinokur, Babich, Zehetmayer}. The ordered phase is characterized by smaller values of $j_{c}$ and the upper critical field $H_{c2}$ than the disordered phase. Let us denote $i = 1$ for the ordered phase and $i = 2$ for the disordered one, which realizes at higher $H$.   
Then the order-disorder transition is described by the Boltzmann function:
\begin{equation}
j_{c}^{\rm peak}(B) = j_{c,1}(B) + \frac{j_{c,2}(B) - j_{c,1}(B)}{1+e^{-(|B|-B_{\rm tr0})/{B_{w0}}}}~~,
\label{jcb_peak}
\end{equation}
where $B_{\rm tr0}$ is the transition field, $B_{w0}$ is the transition width.  
The monotonic function $j_{c,i}(B)$ is given in Appendix.


The near-surface region of superconducting samples does not pin Abrikosov vortices. In works \cite{Clem, Burlachkov} this region is named as the fluxoid-free region. The magnetization of the near-surface region is equilibrium that is a reason of the asymmetry of $M(H)$ loops along the $H$ axis. 
For gross samples an influence of the near-surface region on magnetization loops may be neglected. Then the macroscopic critical current density $J_{c}^{\rm bulk}(H)$ is described by the $j_c(B)$ function with $B=\mu_0H$.
In smaller samples the surface noticeably affects on the macroscopic critical current density $J_c$ and the magnetization that is accounted by the extended critical state model \cite{Chen, Gokhfeld11, Gokhfeld14}.  
Due to avoiding the near-surface region, the macroscopic critical current density $J_c(H)$ depends of the size and the form of samples:
\begin{equation}
J_c(H) = J_{c}^{\rm bulk}(H) \left(1-l_{s}(H)/R \right)^n~~,
\label{Jc_ecsm}
\end{equation}
where $l_s$ is depth of the equilibrium (fluxoid-free) region, $R$ is the radius of the current circulation, $n$ is the index defined by the geometry of the grain ($n = 2$ for a thin plate and $n = 3$ for a cylindrical sample). 
The depth $l_{s}(H)$ inversely correlates with the $j_c(B)$ dependence. 
The peak in the $j_c^{\rm peak}(B)$ dependence is accompanied by a decrease of $l_s$ values such that $l_{s,2}/l_{s,1} \approx j_{c,1} / j_{c,2}$. This relation is observed in asymmetric magnetization loops with the peak effect \cite{gokhfeld_jetp, Altin_nd, Altin_eu, gokhfeld17_1, gokhfeld17}. The depth of the equilibrium region during the order-disorder transition changes as 
\begin{equation}
l_{s}^{\rm peak}(H) = l_{s,1}(H) + \frac{l_{s,2}(H) - l_{s,1}(H)}{1+e^{-(|H|-H_{\rm tr0})/{H_{w0}}}}~~,
\label{ls_peak}
\end{equation}
where $H_{\rm tr0}=B_{\rm tr0}/\mu_0$ and $H_{w0}=B_{w0}/\mu_0$. The phenomenological $l_s(H)$ function is suggested in Appendix.

Inserting functions (\ref{jcb_peak}) and (\ref{ls_peak}) to Eq.~(\ref{Jc_ecsm}) we obtain the second peak in the macroscopic critical current density:
\begin{equation}
J_{c}^{\rm peak}(H) = J_{c,1}(H) + \frac{J_{c,2}(H)-J_{c,1}(H)}{1+e^{-(|H|-H_{\rm tr})/{H_w}}}~~,
\label{Jpeak1}
\end{equation}
with $J_{c,i}(H)=J_{c,i}^{\rm bulk}(H)(1-l_{s,i}(H)/R)^n$, $i=1,2$, $H_w$ is about $H_{w0}$.
The macroscopic critical current density undergoes the transition with the middle at $H = H_{\rm tr}$, which is some higher than $H_{\rm tr0}$. 


Some explanations of the peak effect ground on idea of electromagnetic granularity producing two current systems \cite{Kupfer89,Galuzzi15}. The electromagnetic granularity may emerge due to phase separation in some superconductors.
The phase separation into the insulating and superconducting regions is observed in Ba$_{0.6}$K$_{0.4}$BiO$_3$ superconductor in the range of fields and temperatures overlapping with the peak effect \cite{gokhfeld_jetp}. A network consisting from non-superconducting and superconducting clusters is formed in the sample due to the phase separation. The number and the size of the clusters depend on extrinsic parameters (temperature, transport current, and magnetic field). Upon partial suppression of superconductivity by the magnetic field or the temperature, the volume share of superconducting clusters $P_S$ as well as their size $R$ increase that can be described by the Boltzmann sigmoid function~(\ref{e1}). This is reflected as the second peak in the magnetization loop without the peak in the $j_{c}(B)$ dependence \cite{gokhfeld_jetp}. 
Given the $i$-th state is characterized by $P_{S} = P_{S,i}$ and $R=R_i$, the transition is also described by Eq.~(\ref{Jpeak1}) with $J_{c,i}(H)=P_{S,i}(H) J_{c}^{\rm bulk}(H)\left[1-l_{s}(H)/R_i(H)\right]^n$.

\section{Discussion}
\label{Res}

Equations (\ref{jcb_peak}) and (\ref{ls_peak}) require zero field values of $j_{c,1}$, $l_{s,1}$ and $j_{c,2}$, $l_{s,2}$ to fit experimental magnetization loops. The one pair of the parameters is easy estimated from the width and the asymmetry of magnetization loops. The other pair is connected with an unclear value of the transition width $B_{w0}$. So the ratio $j_{c,2}(0)/j_{c,1}(0)$ is indeterminate. 
There is the value of $j_{c}^{\rm peak}$ at zero field, $j_{c}^{\rm peak}(0)=j_{c0}$, which is independent of $B_{w0}$. The parameter $A =j_{c,2}(0)/j_{c0}$ is easy estimated from magnetization loops. The value of $j_{c0}$ is a combination of $j_{c,1}(0)$ and $j_{c,2}(0)$. To operate with $j_{c0}$ it is convenient to use the sigmoid function, which equals to 0 at $B=0$. We suggest the sigmoid function $S(B) = 1/(1+|B/B_{\rm tr0}|^{-B_{\rm tr0}/B_{w0}})$. This sigmoid function has $S=0$ at $B=0$, $S=0.5$ at $B=B_{\rm tr0}$ and $S(B)$ approaches to 1 at $B \gg B_{\rm tr0}$. Difference between curves computed
with the suggested sigmoid function and curves computed with the Boltzmann sigmoid function is insignificant.
From here, the peak effect due to the order-disorder transition is described by the functions:
\begin{eqnarray}
\label{jcB2}
j_{c}^{\rm peak}(B)= j_{c0}(B) \left(1+ \frac{A-1}{1+\left|B_{\rm tr0}/B \right|^{B_{\rm tr0}/B_{w0}}} \right)~~;\\
\label{lsH2}
l_s^{\rm peak}(H)= l_{s0}(H) \left(1+\frac{\left(\frac{1}{A}-1\right)\left(1-\frac{R}{l_{s0}(H)}\frac{H}{H_{\rm irr}}\right)}{1+\left|H_{\rm tr0}/H \right|^{H_{\rm tr0}/H_{w0}}} \right)~~,
\end{eqnarray}
where $H_{\rm irr}$ is the irreversibility field.
With using these equations, the detailed parametrization of magnetization loops is reached \cite{gokhfeld17}.
The expression for the macroscopic critical current density is obtained from Eq.~(\ref{Jc_ecsm}):
\begin{equation}
\label{JcH2}
J_c^{\rm peak}(H)= J_{c0}(H) \left(1+ \frac{A_J-1}{1+\left|H_{\rm tr}/H \right|^{H_{\rm tr}/H_w}} \right)~~,
\end{equation}
where $A_J=A((R-l_{s0}/A)/(R-l_{s0}))^n$, $H_{\rm tr} \approx H_{\rm tr0}(R-l_{s0}/A)/(R-l_{s0})$ and $H_w=H_{w0}H_{\rm tr}/H_{\rm tr0}$.


\begin{figure}[ht]
\centering
\includegraphics[width=4in]{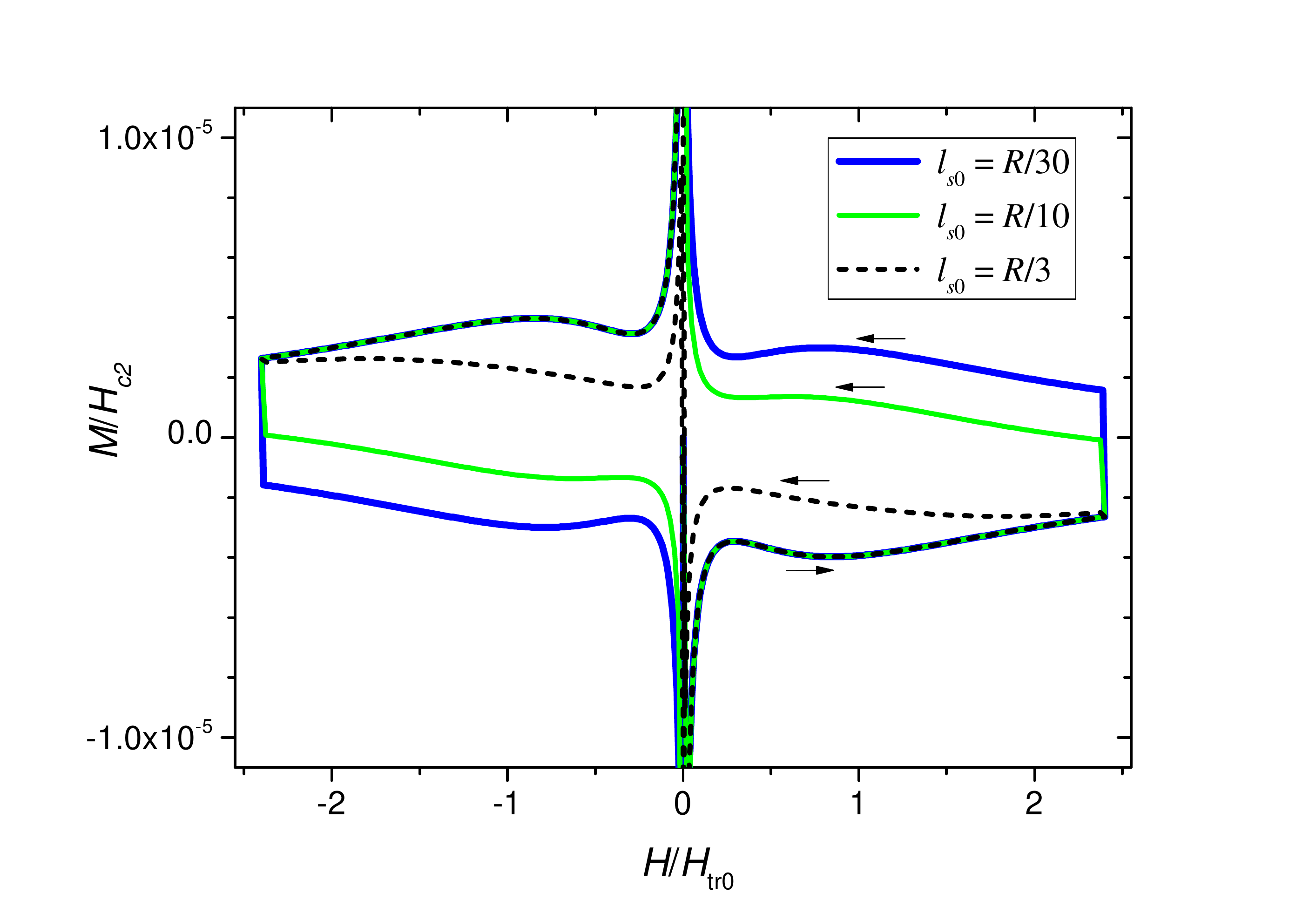}
\includegraphics[width=4in]{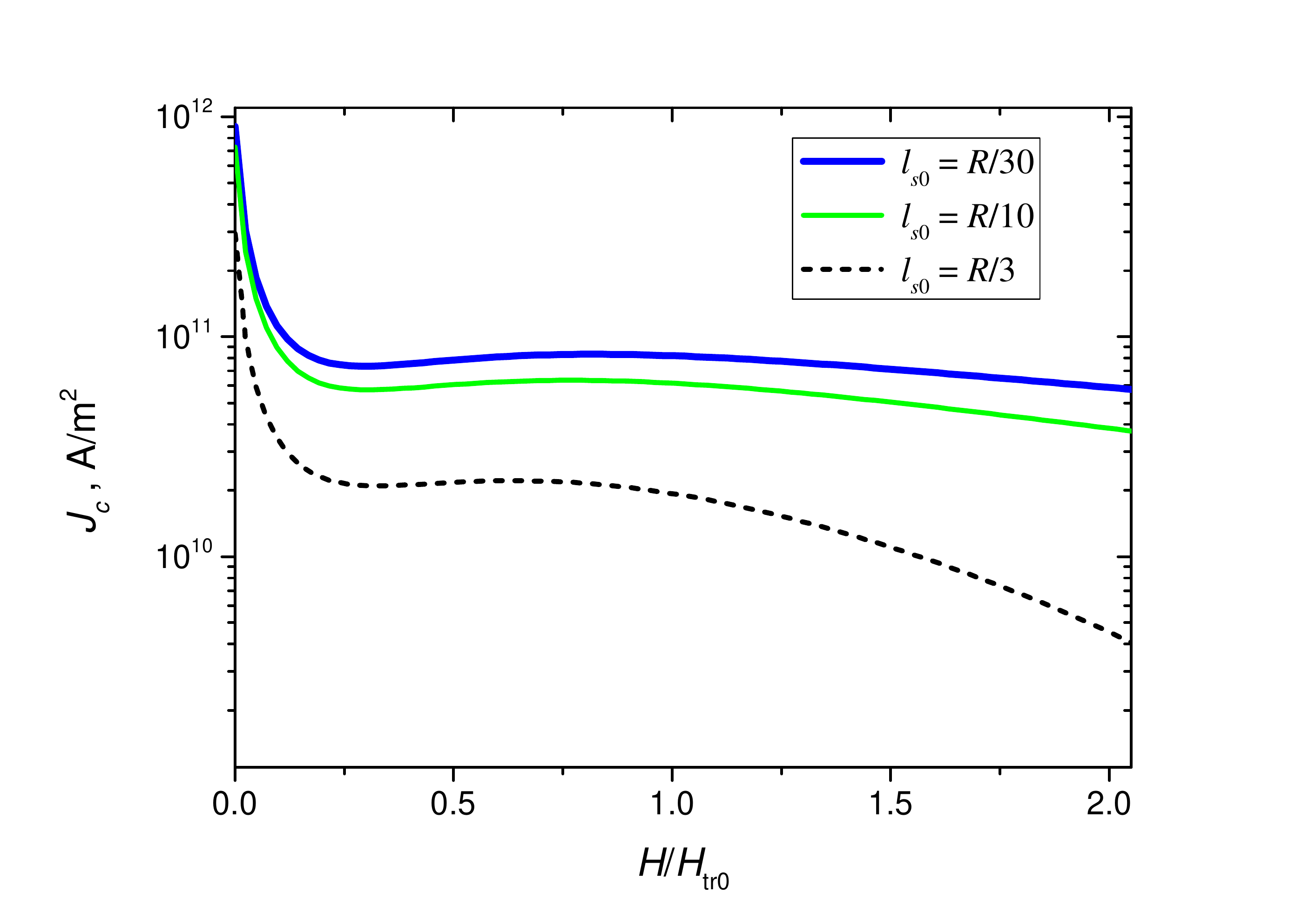}
\caption{Magnetization loops (a) and macroscopic critical current densities (b) computed for various $l_{s0}$. Arrows show the direction of magnetic field change}
\label{fig_mh_ls}
\end{figure}

Fig.~\ref{fig_mh_ls}a shows the magnetization loops computed with different depths of the surface equilibrium region for the case of the order-disorder transition. For all the plotted loops the peak height $A$ equals to 15. All the loops are computed with the same values of $H_{c2}$ and $H_{\rm tr0}=0.01 H_{c2}$, $H_{w0} = 0.5 H_{\rm tr0}$. The values of $H_{\rm irr}$ depend on the $l_{s0}/R$ ratio: $H_{\rm irr}=0.302 H_{c2}$ for $l_{s0}= R/30$, $H_{\rm irr}=0.100 H_{c2}$ for $l_{s0}= R/10$, and $H_{\rm irr}=0.029 H_{c2}$ for $l_{s0}= R/3$.
It is seen that  the second peak is presented on both a magnetization branch for the growing magnetic field and a branch for the reversed magnetic field. Magnetization loops with the higher $l_{s0}/R$ ratio are more asymmetric and have the less pronounced second peak in the branch for the reversed magnetic field.
The macroscopic critical current density corresponding to the magnetization loops in Fig.~\ref{fig_mh_ls}a is presented in Fig.~\ref{fig_mh_ls}b. The $J_c(H)$ dependencies decrease faster for the higher $l_{s0}/R$ ratio. The observed peak in the $J_c(H)$ dependencies moves to lower $H$ as the $l_{s0}/R$ ratio increases.

The position of the second peak depends on $T$ and $R$ \cite{gokhfeld17_1, gokhfeld17, Kalisky, Krelaus99}. For thin superconducting films, small samples, and polycrystalline samples consisting from small grains the second peak locates near zero $H$ and may be unobservable.
Also the second peak position is expected to be influenced by the angle between the magnetic field direction and crystallographic planes of an anisotropic superconductor.  Shift of $H_{\rm peak}$ within a required field range may be desirable for some applications.

\section{Conclusions}

The second peak in $M(H)$ loops is resulted from the magnetic transition from the state with lower $J_{c0}$ values to the state with the higher ones. This takes place due to the order-disorder transition of the vortex lattice or to the phase separation. The peak appearance request such a change of some parameters (t.g. $j_c$, $l_s$, $P_S$, $R$), that their evolutions are described by the Boltzmann sigmoid function. The peak in the $J_c(H)$ dependence can occur without the corresponding peak in the local critical current density $j_c(B)$.

The magnetization loops with the fishtail were computed by using the extended critical state model and a sigmoid function as the source of the peak. The simplicity of equations (\ref{jcB2}), (\ref{lsH2}) and (\ref{JcH2}) make them suitable for parametrization 
of magnetization loops.
The presented approach reproduces various magnetization loops with the second peak. Recently magnetization loops of Y$_{1-x}$Nd$_x$Ba$_2$Cu$_3$O$_{7-\delta}$ $(x = 0.02, 0.11, 0.25)$ superconductors were successfully described \cite{gokhfeld17_1, gokhfeld17}.

\section{Appendix}

The dependence of the local critical current density $j_c$ on the inner magnetic field $B$ is described by a decreasing function $j_c(B)$. The Kim \cite{Kim62}, a power \cite{Irie67} and an exponential \cite{Chen} model are usually used. 
We support the following generalized $j_c(B)$ dependence \cite{SST17}:
\begin{equation}
\label{jcb}
j_c(B) = j_{c0}\frac{1-|B/B_{c2}|^\alpha}{1+|B/B_{0}|^\alpha}~~,
\end{equation}
where $B_{c2}= \mu_0 H_{c2}$, $\alpha$ is positive dimensionless coefficient. 
This function gives better agreement with experimental dependencies in field range from 0 to $H_{c2}$ than the earlier generalized dependence \cite{Kumar89}. 

The simple phenomenological $l_s(H)$ dependence is written as
\begin{equation}
l_s(H) = l_{s0} \left(1 + |H|/H_{1}\right)~~,
\label{ls_Hc2}
\end{equation}
where $H_1$ is the increasing rate. 
The magnetization loops becomes reversible in $H$ higher than the irreversibility field $H_{\rm irr}$. So the $l_s(H)$ dependence increases from $l_{s0}$ at $H = 0$ to $R$ at $H = H_{\rm irr}$. Eq.~(\ref{ls_Hc2}) can be rewritten as
\begin{equation}
l_s(H,T) = l_{s0} + (R - l_{s0}) |H|/H_{\rm irr}~~.
\label{ls_Hirr}
\end{equation}
As distinct from $H_1$, the value of $H_{\rm irr}$ depends on the size $R$. 

Expressing $J_c(H)$ at $H = 0$ as $J_{c0} = j_{c0} (1-l_{s0}/R)^n$, one obtains the magnetic field dependence of the macroscopic critical current density:
\begin{equation}
J_c(H)=J_{\rm c0} \frac{1-\left|H/ H_{c2}\right|^\alpha}{1+\left| H/H_{0} \right|^\alpha} \left(1-|H/H_{\rm irr}|\right)^n~~.
\label{Jch}
\end{equation}
A scaling of pinning force at different temperatures is resulted from this equation \cite{SST17}. Eq. (\ref{Jch}) successfully describes $J_c(H)$ dependencies for most superconductors without the peak effect.

\end{document}